\documentclass{article}
\usepackage[utf8]{inputenc}
\usepackage{amsmath,amsthm,amssymb,amsfonts}
\usepackage{graphicx}
\usepackage{todonotes}
\usepackage{hyperref}
\usepackage{tikz} 

\font\header=cmssdc10 at 20pt

\def\1{{\bf 1}}

\begin{document}

{\header Evolutionary paths under catastrophes }

\vskip1cm

Rinaldo B. Schinazi
\let\thefootnote\relax\footnote{Department of Mathematics, University of Colorado, Colorado Springs, CO 80933-7150, USA;
rinaldo.schinazi@uccs.edu}

\vskip1cm

{\bf Abstract.} We introduce a model to study the impact of catastrophes  on evolutionary paths. If we do not allow catastrophes the number of changes in the maximum fitness of a population grows logarithmically with respect to time. Allowing catastrophes (no matter how rare) yields a drastically different behavior. When catastrophes are possible the number of changes in the maximum fitness of the population grows linearly with time. Moreover, the evolutionary paths are a lot less predictable when catastrophes are possible. Our results can be seen as supporting the hypothesis that catastrophes speed up evolution by disrupting dominant species and creating space for new species to emerge and evolve.

\vskip1cm


\vskip1cm

{\header 1 The model}

\vskip1cm

We think of a catastrophe as an 
 event that causes a major change in the environment to the extent that it renders a species accumulated adaptation essentially useless, even if not many individuals die. Suppose one specialized aspect of a species has evolved to be suitable to a particular ecological niche, and then that niche disappears (e.g. due to climate change, or habitat loss, or extinction of prey). Then this particular aspect of that species fitness becomes irrelevant and has to start over as it begins adapting to a new reality. This view of catastrophes seems to support the hypothesis that evolution speeds up in a rapidly changing environment.
Next we introduce a probability model to test this hypothesis.

Let $p$ be a real number in $[0,1]$.  Consider a sequence $(V_n)_{n\geq 0}$ of independent identically distributed (i.i.d. in short) random variables. Assume that the common distribution of the $V_n$ is continuous. For instance, we
 will use a mean 1 exponential distribution in the simulations.  At every discrete time $n\geq 1$ there is exactly one birth and the fitness $V_n$ is assigned to the new individual. We define the process $(M_n)_{n\geq 0}$ as follows.  Let $M_0=V_0$ and for $n\geq 0$,  there are two possibilities.  
\begin{itemize}
    \item With probability $p$ there is no catastrophe at time $n$ and $M_{n+1}=\max(M_n,V_{n+1})$.
    \item With probability $1-p$ there is a catastrophe and $M_{n+1}=V_{n+1}$.
\end{itemize}
  We interpret $M_n$ as being the maximum fitness of the population at time $n$. In between catastrophes $(M_n)$ can only go up or stay put depending on its current value and the random sampling of the new birth.
  On the other hand if there is a catastrophe at time $n$ then the line of evolution is destroyed and a new line starts anew at time $n$.

 Let $X_t$ be the number of $s$'s for which $0\leq s\leq t$ and $M_s$ is different from all $M_u$ for $0\leq u<s$. That is, $X_t$ counts the number of times the process $(M_n)$ takes a new value up to time $t$. 
 The process $(X_t)$ can be thought of as a measure of evolution speed. We will show that without catastrophes $X_t$ increases as $\ln t$ as $t\to\infty$  while it increases linearly when catastrophes are possible.

\vskip1cm

{\header 2 No catastrophes}

\vskip1cm

Assume that $p=1$. That is, catastrophes do not happen. Then, at all times $n\geq 0$, $M_n$ can be written as
$$M_n=\max(V_0,V_1,\dots,V_n).$$

In this case $X_t$ is the number of times $(M_n)$ has gone up for $n\leq t$. In the probability literature $X_t$ is known as the number of records up to time $t$. A classical result is,

\medskip

$\bullet$ For $p=1$, almost surely,
$$\lim_{t\to\infty}\frac{1}{\ln t}X_t=1.$$

\medskip

Thus, without catastrophes the number of distinct values that $(M_n)$ takes grows only logarithmically with time. The key observation for the analysis of the number of records is the following. Let $i\geq 0$ and let $A_i$ be the event
$$V_i=\max\{V_0,V_1,\dots,V_i\}.$$
Then, by symmetry $P(A_i)=\frac{1}{i+1}.$ Since
$$X_t=\sum_{i=0}^t 1_{A_i},$$
we get 
$$E(X_t)=\sum_{i=0}^t \frac{1}{i+1}\sim \ln t,$$
as $t$ goes to infinity.
For more details as well as other results on the number of records see for instance Port (1994). Observe that the results about the process $(X_t)$ do not depend on the specific (continuous) distribution of $V_n$.

\vskip1cm

{\header 3 Catastrophes}

\vskip1cm

Consider the case $p<1$. That is, catastrophes happen with probability $1-p>0$ at every unit time. We will prove the following result.

\medskip

{\bf Proposition 1. }{\sl  For $0< p<1$, almost surely,
$$\lim_{t\to\infty}\frac{1}{t}X_t=-\frac{(1-p)}{p}\ln(1-p).$$
}

\medskip

Hence,  the number of distinct values that $(M_n)$ takes grows linearly with time and we have an exact expression for the slope. This is in sharp contrast with the model with no catastrophes.

\vskip1cm

{\header 4 Discussion}

\vskip1cm

There is a number of population biology models with catastrophes, see  Brockwell (1986) and Neuts (1994) for instance. Those models study the fluctuations of the size of a population. Our focus is on evolutionary paths instead.

Closer to our point of view are the probability models introduced to model evolutionary paths. In these models the evolutionary path gets stuck after a few steps because the so-called fitness landscape is fixed and a transition can only occur if it increases the fitness, see Gillespie (1983), Kaufman and Levin (1987) and Hegarty and Martinson (2014). So once the path hits a local maximum in the fitness landscape it is stuck there forever. No such thing can happen in our model. Maxima are never permanent, see also Schinazi (2019). 

The main difference with previous models for evolutionary paths is our introduction of catastrophes. In our model for any value of $p<1$ (i.e. catastrophes happen with probability $1-p>0$) we see a drastic change in the behavior of evolutionary paths as compared to $p=1$ (i.e. no catastrophes), see Figures 1 and 2. In particular the number of jumps in a path grows linearly  for any $p<1$ and only logarithmically for $p=1$. We interpret the number of jumps in a path as a measure of evolution speed. Hence, our results can be seen as supporting the hypothesis that catastrophes speed up evolution by disrupting dominant species and creating space for new species to emerge and evolve.

\vskip1cm

{\header 5 Proof of Proposition 1}

\vskip1cm

We first introduce some notation. Recall that we have an i.i.d. sequence of fitnesses $V_0,V_1,\dots$ where $V_t$ is the fitness of the individual born at time $t$. For $0\leq s<t$, let $R(s,t)$ be the number of records in the sequence $(V_s,V_1,\dots,V_t)$. That is, $R(s,t)$ is the number of $u's$ such that $s\leq u\leq t$ and
$$V_u=\max\{V_s,\dots,V_u\}.$$
Observe that $R(s,t)$ ignores catastrophes. It counts records between two catastrophes.
Note also that by our definition a record always happens at the initial time $s$.

Assume that $0<p<1$. Let $T_0=0$ and for $i\geq 1$ let $T_i$ be the time of the $i$-th catastrophe. For $t\geq 0$, let 
$$N_t=\max\{n\geq 0:T_n\leq t\}.$$
That is, $N_t$ is the number of catastrophes up to time $t$. 

There are two ways for $M_t$ to take a value that has not been taken before by $(M_n)$.

\begin{itemize}
    \item If there is a catastrophe at time $t$ then $M_t$ takes a value never taken before in $(M_n)$. This is so because we assume that the fitness distribution is continuous.
    \item If $s<t<u$ where successive catastrophes happened at times $s$ and $u$ then $(M_n)$ takes a new value at time $t$ if and only if there is a record at time t for the sequence of fitnesses that started at time $s$.
\end{itemize}

 Using the two observations above the number of distinct values $X_t$ that $(M_n)$ has taken up to time $t$ can be written as
$$X_t=\sum_{i=1}^{N_t}R(T_{i-1},T_i-1)+R(T_{N_t},t).\leqno (1)$$

Let $t\geq 0$, $N_t$ has a binomial distribution with parameters $1-p$ and $t$. By the Law of Large Numbers,  almost surely
$$\lim_{t\to\infty}\frac{N_t}{t}=1-p.\leqno (2)$$

Since the sequence $(T_i-T_{i-1})_{i\geq 1}$ is i.i.d.  so is the sequence $\left(R(T_{i-1},T_i-1)\right)_{i\geq 1}$. By
the Law of Large Numbers, almost surely,
$$\lim_{t\to\infty} \frac{1}{N_t}\sum_{i=1}^{N_t}R(T_{i-1},T_i-1)=E(R(0,T_1)).$$
Using (2),
$$\lim_{t\to\infty} \frac{1}{t}\sum_{i=1}^{N_t}R(T_{i-1},T_i-1)=(1-p)E(R(0,T_1-1)).\leqno (3)$$

We now turn to $R(T_{N_t},t)$. 
Observe that
$$T_{N_t}=\sum_{i=1}^{N_t} (T_i-T_{i-1}).$$
Since the sequence $(T_i-T_{i-1})_{i\geq 1}$ is i.i.d. with a geometric $1-p$ distribution the Law of Large Numbers applies. Hence,
$$\lim_{t\to\infty}\frac{T_{N_t}}{N_t}=E(T_1)=\frac{1}{1-p}.$$

Writing,
$$\frac{T_{N_t+1}}{T_{N_t}}=\frac{T_{N_t+1}}{N_t+1}\frac{N_t}{T_{N_t}}\frac{N_t+1}{N_t},$$
we see that,
$$\lim_{t\to\infty}\frac{T_{N_t+1}}{T_{N_t}}=1.$$
Thus,
$$\lim_{t\to\infty} \frac{T_{N_t}}{t}=1.$$

The number of records between times $s$ and $u$ is at most $u-s+1$. Hence,
$$R(T_{N_t},t)\leq t-T_{N_t}+1.$$
Therefore,
$$\lim_{t\to\infty}\frac{1}{t}R(T_{N_t},t)=0.\leqno (4)$$
Using (3) and (4) in (1) we get
$$\lim_{t\to\infty}\frac{X_t}{t}=(1-p)E(R(0,T_1-1)).$$
To finish the proof we need to compute $E(R(0,T_1-1))$.
Note that
\begin{align*}
R(0,T_1-1)=&\sum_{i=0}^{T_1-1} 1_{A_i}\\
=&\sum_{i=0}^\infty 1_{B_i},
\end{align*}
where $B_i=\{T_1>i\}\cap A_i$. Recall that
$A_i$ be the event
$$V_i=\max\{V_0,V_1,\dots,V_i\}.$$
Note that $A_i$ is a function of $\{V_0,\dots,V_i\}$ while $T_1$ depends on i.i.d. Bernoulli random variables with parameter $p$ that are independent of the sequence $(V_n)$. Hence, $\{T_1>i\}$ and $A_i$ are independent events. Moreover, $T_1$ has a geometric distribution with parameter $1-p$. Therefore,
\begin{align*}
 E(R(0,T_1-1))=&\sum_{i=0}^\infty P(A_i)P(T_1>i)\\
 =&\sum_{i=0}^\infty p^i\frac{1}{i+1}\\
 =&-\frac{1}{p}\ln(1-p).
\end{align*}

Hence,
$$\lim_{t\to\infty}\frac{X_t}{t}=-\frac{1-p}{p}\ln(1-p).$$

\vskip1cm

{\header References}

\vskip1cm

P.J. Brockwell (1986).
The extinction time of a general birth and death process with catastrophes.
Journal of Applied  Probability 23, 851-858.

J.H. Gillespie (1983) A simple stochastic gene substitution model. Theoretical  Population Biology 23, 202-215.

P. Hegarty and S. Martinsson (2014) On the existence of accessible paths in various models of fitness landscapes. Annals of Applied Probability 24, 1375-1395.

S. Kaufman and S. Levin (1987) Towards a general theory of adaptive walks in rugged landscapes. Journal of Theoretical Biology 128, 11-45.

R.E. Lenski and M. Travisano (1994) Dynamics of adaptation and diversification: a 10,000-generation experiment with bacterial populations. Proceedings of the National Academy of Sciences (U.S.A.) 91, 6808-6814.

M.F. Neuts (1994). An interesting random walk on the non-negative integers.
Journal of Applied Probability 31, 48-58.

S.C. Port (1994) {\sl Theoretical probability for applications}, Wiley.

R.B. Schinazi (2019) Can evolution paths be explained by chance alone? Journal of Theoretical Biology 465, 65-67.

\bigskip

{\bf Acknowledgements.} We thank two anonymous referees whose constructive criticism was quite helpful in improving the paper.

\begin{figure}[ht]
	\includegraphics[width=10cm]{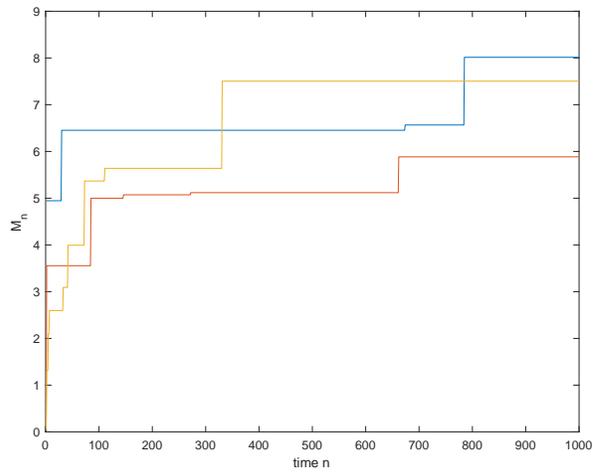}
	\caption{ These are three independent simulations of the paths of the process $(M_n)$ for $0\leq n\leq 1000$. The fitness distribution is an exponential mean 1 distribution. We take $p=1$ (i.e. no catastrophes). The shapes of the three paths are similar and somewhat parallel. This is consistent with what was observed by Lenski and Travisano (1994) in their bacteria experiments. The number of jumps for each path is small and all jumps are upwards.}
	\end{figure}

\begin{figure}[ht]
	\includegraphics[width=10cm]{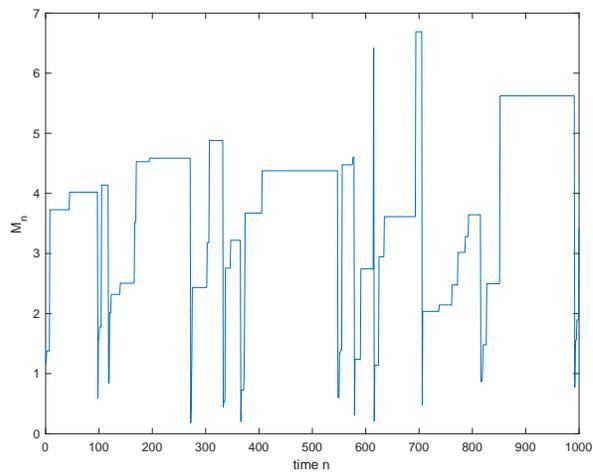}
	\caption{ This is a simulation of $(M_n)$ for $0\leq n\leq 1000$. The fitness distribution is an exponential mean 1 distribution. We take $p=0.99$. As compared to Figure 1 the path is a lot less predictable. Sections of the path between two successive catastrophes can be very different in length and shape. The number of jumps in the path is large. Jumps can be upwards or downwards.}
	\end{figure}

\begin{figure}[ht]
	\includegraphics[width=10cm]{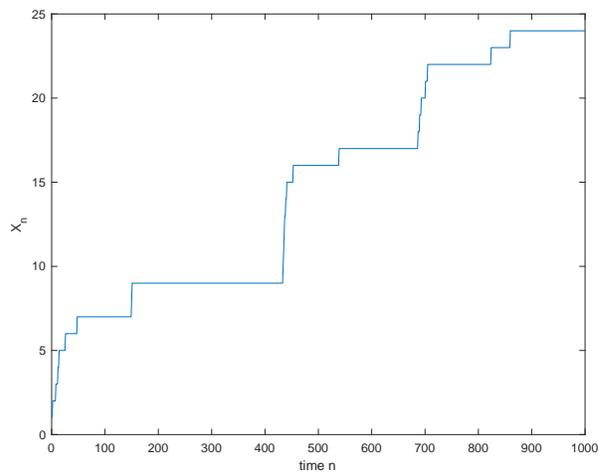}
	\caption{This is a simulation of $(X_n)$ for $0\leq n\leq 1000$. The fitness distribution is an exponential mean 1 distribution. We take $p=0.99$. As predicted by Proposition 1 $X_n$ grows linearly with time. The times at which jumps occur are quite random. This is consistent with the simulation in Figure 2.}
\end{figure}

\end{document}